\documentstyle[psfig]{article} \catcode`\@=11
\newcommand{\bbox}[1]{{\mbox{\boldmath$#1$}}}
\newcommand{\bp}{{\bbox{p}}}
\newcommand{\bq}{{\bbox{q}}}
\newcommand{\ds}{\displaystyle}
\newcommand{\ba}{{\bbox{\alpha}}}
\newcommand{\hba}{{\hat\ba}}
\newcommand{\bb}{{\bbox{\beta}}}
\newcommand{\be}{\begin{equation}}
\newcommand{\ee}{\end{equation}}
\newcommand{\bI}{{\bbox{I}}}
\newcommand{\bth}{{\bbox{\theta}}}
\newcommand{\bw}{{\bbox{\omega}}}
\newcommand{\bnu}{{\bbox{\nu}}}
\newcommand{\rd}{{\rm d}}
\newcommand{\ddt}{{\frac{\rd}{\rd t}}}
\newcommand{\dt}{{\rm d}t\,}
\newcommand{\meta}{{\hat {\cal R}}}
\newcommand{\vac}{{|0\rangle}}
\newcommand{\by}{{\bbox{y}}}
\newcommand{\bx}{{\bbox{x}}}
\newcommand{\T}{{\cal T}}

    \def\section{\@startsection{section}{1}{\z@}
    {-3.5ex plus -1ex minus -.5ex}{1.5ex plus.3ex}{\bf }}
    \def\subsection{\@startsection{subsection}{1}{\z@}
    {-3.5ex plus-1ex minus-.5ex}{1.5ex plus.3ex}{\em }} 
    
\begin{document}
\mbox{} \vspace{.75cm} \newline{\Large\bf
Semiclassical trace formulae using
coherent states
    }\vspace{.4cm}\newline{\bf   
Bernhard Mehlig$^{1,}$\footnote[2]{\tt mehlig@physik.uni-freiburg.de}
and Michael Wilkinson$^3$
    }\vspace{.4cm}\newline\small
$^1$Theoretische Quantendynamik, Fakult\"at f\"ur Physik,
Universit\"at Freiburg, D-79104 Freiburg\\ 
$^3$Faculty of Mathematics
and Computing, The Open University, Milton Keynes, MK7 6AA,
Bucks, England.
    \vspace{.2cm}\newline 
Received 4 May 2000, revised 28 September 2000, accepted 1 November 2000 by U. Eckern
    \vspace{.4cm}\newline\begin{minipage}[h]{\textwidth}\baselineskip=10pt
    {\bf  Abstract.}
We derive semiclassical trace formulae including Gutzwiller's
trace formula using coherent states.
This formulation has several advantages over
the usual coordinate-space formulation.
Using a coherent-state basis makes it immediately
obvious that classical periodic orbits make separate
contributions to the trace of the quantum-mechanical
time evolution operator.
In addition, our approach
is manifestly canonically invariant at all
stages, and leads to the simplest possible
derivation of Gutzwiller's formula.
    \end{minipage}\vspace{.4cm} \newline {\bf  Keywords:}
semiclassical approximation, trace formulae, coherent states
\newline {\bf  PACS:} {05.45.+b}
    \newline\vspace{.2cm} \normalsize
\section{Introduction}
\label{sec:sec1}
Gutzwiller's trace formula \cite {gut67,gut71,gut90}
relates the density of states of a quantum system to periodic
orbits of the corresponding classical Hamiltonian. In its
usual form it applies to systems with a discrete quantum
spectrum, and isolated, unstable periodic orbits. In the general
case it is not an exact relation. The density
of states $\rho(E)$ may be expressed in terms of the trace
of the quantum-mechanical time evolution operator
\be
\label{eq:eqno(1.1)}
\rho(E) = \sum_n\delta(E-E_n)=
\int_{-\infty}^\infty 
\frac{\dt}{2\pi\hbar} \exp({\rm i}Et/\hbar){\rm Tr}[\hat U_t] 
\ee
where $E_n$ are the eigenvalues and $\hat U_t=\exp(-{\rm i}\hat Ht/\hbar)$
is the quantum-mechanical time evolution operator. 
The idea is to approximate the trace
by a formula valid in the semiclassical ($\hbar \to 0$) limit.
Gutzwiller has also provided 
a discussion of some precursors of his expression in the mathematical 
literature \cite{gut90}, including an exact formula due to
Selberg (discussed in \cite{bal87}) which  applies to the Schr\"odinger
equation on closed manifolds of constant negative curvature.
Gutzwiller's formula has been used as a basis for many 
investigations in semiclassical quantum mechanics,
with applications in atomic \cite{win87}, molecular and mesoscopic physics
\cite{montambaux}. We will discuss some of its extensions
in due course. It is certainly a powerful and fundamental
element in semiclassical quantum theory, and it is desirable 
to have a clear and direct derivation.

The original derivation by Gutzwiller 
is undeniably far more
complicated than the final result, and it also has the disadvantage
that it 
uses the coordinate representation of the quantum-mechanical
time evolution operator, whereas the final 
answer is expressed in a canonically
invariant form. The approach described here uses a coherent-state 
basis to evaluate the trace. The coherent states 
$\vert \ba \rangle$ will be defined later. They are labeled
by a phase-space point $\ba   =({\bq},{\bp})$ --- $\bq$
and $\bp$ are $d$-dimensional
coordinate and momentum vectors respectively --- and may
be thought of as wave packets positioned at the phase-space point
$\ba$. The trace of any operator may be expressed as an 
integral over coherent states. For example, the trace of
$\hat U_t$ may be written as
\be
\label{eq:eqno(1.2)}
{\rm Tr}[\hat U_t ]=\int\!
\frac{\rd\ba}{(2\pi \hbar)^d}\,
\langle \ba    \vert \hat U_t\vert \ba    \rangle \ .
\ee
The semiclassical approximation for the state resulting from 
the action of the time evolution operator $\hat U_t $ on the coherent state 
$\vert \ba \rangle$ is clearly expected to be a wave packet
with phase-space coordinates $\ba_t$, which are the 
result of mapping the phase-point $\ba$ for time $t$ under
the action of Hamilton's equations. The contribution to the integral
(\ref{eq:eqno(1.2)}) from the phase point $\ba$ is clearly negligible,
unless $\ba $ is very close to a periodic orbit of period close to 
$t$. The fact that the density of states may be expressed in terms of
periodic orbits is thus immediately obvious when the coherent state
basis is used. This point was made some time ago in a paper
by one of us \cite{wil87}. 

It is less obvious how the precise
form of Gutzwiller's trace formula may be obtained using
coherent states. 
A derivation based upon coherent states has previously been 
given by Combescure {\em et al} \cite{com97}, but we will
argue shortly that our method has the advantage of being
canonically invariant, as well as being considerably simpler.
Our calculation builds upon the semiclassical evolution
of coherent states as described in \cite{lit86}.
Translation of wave packets in phase space
is effected by so-called {\em Weyl-Heisenberg operators},
$\hat \T(\ba)$.
Wave packets are not only moved around in phase space
but, in general, also deformed under quantum evolution.
This deformation is described by a symplectic transformation
represented by a matrix $\tilde S$. A representation
of symplectic transformations on the quantum-mechanical
Hilbert space  is achieved by {\em metaplectic operators}
$\meta(\tilde S)$ 
(they actually form a double covering of the symplectic group). 
In \cite{lit86} the metaplectic operators are 
expressed in terms of coordinate-space matrix elements,
and a similar coordinate space representation was used 
by Combescure {\em et al}. Our calculation uses the Weyl
representation of metaplectic operators, partially 
developed in \cite{wil98}, in which 
an operator $\hat A$ is represented as an integral over
Weyl-Heisenberg operators $\hat \T(\ba)$, with weight
$A(\ba)$. 
This representation  
gives a canonically invariant formulation, without
reference to any coordinate space representations.

Our derivation of Gutzwiller's formula
is based upon a simple formula for the trace of a metaplectic 
operator
\be
\label{eq:eqno(1.3)}
{\rm Tr}[\meta(\tilde S)]={\exp({\rm i}\pi \nu /2)
\over{\sqrt{\vert {\rm det}(\tilde S-\tilde I)\vert}}}
\ee
(where $\nu $ is an integer index).
It uses the fact that the periodic-orbit
contributions to ${\rm Tr}[\hat U_t]$ are proportional
to $\exp({\rm i} R_{\rm p}/\hbar) {\rm Tr}[\meta(\tilde M_{\rm p})]$
where $R_{\rm p}$ is Hamilton's action along the periodic orbit
and $\tilde M_{\rm p}$ is the corresponding stability matrix.
The use of this expression leads to a derivation which is
arguably the simplest possible, 
since none of the intermediate stages involve expressions 
which are significantly more complicated than the final result. 
All of the stages are canonically invariant.

An advantage of finding a simplified derivation is that
extensions and variations of the original expression are
easily obtained. 
Gutzwiller's trace formula has been extended to give an expression 
for the density of states of integrable systems
\cite{ber77}. It has also been extended
by one of us to yield information about diagonal \cite{wil88}
and off-diagonal \cite{wil87} matrix elements of operators.
We will also give very direct derivations of the
latter two formulae, and discuss the integrable case 
briefly in the concluding section.

The remainder of this paper
is organized  as follows.
Section \ref{sec:sec2} deals with some
definitions and notation. 
Section \ref{sec:sec3}
discusses a semiclassical
approximation 
using the Weyl representation of
metaplectic operators. This representation
of metaplectic operators 
has not previously been developed in detail. 
It allows us to work
entirely in phase space, thus simplifying
the discussion considerably. In section \ref{sec:sec4} we show how
to derive a trace formula
for the density of eigenphases
of a Floquet matrix corresponding to
a chaotic quantum map. While this
example has limited physical applications, it serves
as a simple illustrative example.
In section \ref{sec:sec4.2} we
derive the Gutzwiller trace formula for chaotic
flows. The discussion is parallel
to that of section \ref{sec:sec4},
with the complication
that the marginal direction
along the flow has to be treated
separately. Section \ref{sec:sec4.3}
considers more general trace formulae
for chaotic systems. Section
\ref{sec:sec5} contains 
a discussion of the results.
\section{Definitions and notation}
\label{sec:sec2}
\subsection{Traces}
We consider a closed quantum system described
by a time-independent Hamiltonian $\hat H$,
for which the solution of the stationary Schr\"odinger 
equation gives rise
to discrete energy levels $E_j$
and eigenfunctions $|\psi_j\rangle$,
$\hat H |\psi_j\rangle = E_j\,|\psi_j\rangle$.
 The {\em density of states}
is defined as
\be
\label{eq:eqno(2.1)}
\rho (E) \equiv \sum_j\delta(E-E_j) \,.
\ee
In the following it will be shown
how to obtain semiclassical estimates of
this and more general densities such as
\begin{eqnarray}
\label{eq:eqno(2.2)}
\rho_A(E)  &=& \sum_j\langle\psi_j|\hat A|\psi_j\rangle\,\delta(E-E_j)\,,\\
\label{eq:eqno(2.3)}
S_A(E,\hbar\omega) &= &
\sum_{j,l} |\langle\psi_j|\hat A|\psi_l\rangle|^2\,
\delta(\hbar\omega-E_k+E_j)\, \delta(E-E_j)\,.
\end{eqnarray}
Here, $\hat A$ is an observable with a classical limit.
The three densities $\rho (E), \rho_A(E)$ and $S_A(E,\hbar\omega)$ 
may be written as traces over
the quantum-mechanical time evolution operator
\be
\label{eq:eqno(2.4)}
\hat U_t \equiv
{\rm exp}\big(-{\rm i}\hat H t/\hbar\big)\,.
\ee
One has
\begin{eqnarray}
\label{eq:eqno(2.5)}
\label{eq:dos}
\rho (E) &=& \int_{-\infty}^\infty\!\frac{\rd t}{2\pi\hbar}\, 
\exp ({\rm i}Et/\hbar)\,
\mbox{Tr}\,[\hat  U_t]\,,\\
\label{eq:eqno(2.6)}
\rho_A(E) &=& \int_{-\infty}^\infty\!\frac{\rd t}{2\pi\hbar}\, 
\exp({\rm i} Et/\hbar)\,
\mbox{Tr}\,[\hat A\,\hat  U_t]\,,\\
\label{eq:eqno(2.7)}
S_A(E,\hbar\omega) &=& \int_{-\infty}^\infty\!\frac{\rd t}{2\pi\hbar}\,
\exp({\rm i}Et/\hbar)\, \int_{-\infty}^\infty\!
\frac{\rd s}{2\pi\hbar}\,
\exp({\rm i}\omega s)
\mbox{Tr}\,[\hat A\,\hat U_s\, \hat A\, \hat U_{t-s}]\,.
\end{eqnarray}
In the remainder of this article, semiclassical
expressions for the traces occurring in Eqs. (\ref{eq:eqno(2.5)})
to (\ref{eq:eqno(2.7)}) will be sought.

We will also consider explicitly time-dependent systems,
subject to a driving force with period $T$. In this
case the {\em density of eigenphases} $\theta_j$ of the Floquet operator
$\hat{U}_T$ will be calculated. 
The Floquet operator
maps solutions $|\psi_t\rangle$ of the time-dependent
Schr\"odinger equation according
to $|\psi_{t+k\,T}\rangle = \hat{U}_T^k\,|\psi_t\rangle$
where $k$ is an integer.
$\hat{U}_T$ is a unitary operator and 
the density of its eigenphases is defined as
\be
\label{eq:eqno(2.8)}
\rho (\theta) = \sum_{l=-\infty}^\infty\sum_{j =  1}^N 
\delta(\theta-\theta_j-2\pi l)\,.
\ee
This may be written as
\be
\label{eq:eqno(2.9)}
\rho (\theta) = \frac{1}{2\pi\hbar}
\sum_{n= -\infty}^\infty 
\sum_{j= 1}^N 
{\rm e}^{\displaystyle {\rm i} n(\theta-\theta_j)}
\ee
using the Poisson summation formula. Thus
\be
\label{eq:eqno(2.10)}
\rho (\theta) = \frac{1}{2\pi\hbar}
\sum_{n=-\infty}^\infty \mbox{Tr}\,[\hat U_T^n]\,  
{\rm e}^{\displaystyle {\rm i} n\theta}
= \frac{N}{2\pi\hbar} +\frac{1}{\pi\hbar}\mbox{Re} 
\sum_{n>0} \mbox{Tr}\,[\hat U_T^n]\,  
{\rm e}^{\displaystyle {\rm i} n\theta}\,.
\ee
In this case, we seek a semiclassical expression
for $\mbox{Tr}\,[\hat U_T^n]$ for $n > 0$.
Eq. (\ref{eq:eqno(2.10)}) may be viewed
as a discrete analogue of Eq. (\ref{eq:dos}).

\subsection{Classical dynamics}
We consider a possibly time-dependent Hamiltonian
flow defined by the Hamilton function $H(\bq,\bp;t)$ with
$d$ degrees of freedom.
Here $\ba  = (\bq,\bp) = (q_1,\ldots,q_d,p_1,\ldots,p_d)$ 
are phase-space coordinates, $\ba_t$ is a family
of phase-space points representing a trajectory labeled
by time $t$.  The flow is governed by {\em Hamilton's equations}
\be
\label{eq:eqno(2.11)}
\dot\ba_t = \tilde J \frac{\partial H}{\partial \ba_t}
\ee
with
\be
\label{eq:eqno(2.12)}
\tilde J = 
\left(\begin{array}{cc} \tilde 0 & \tilde I\\ -\tilde I & \tilde 0 
\end{array}\right)\,.
\ee
The separation $\delta\ba_t$ of two nearby trajectories
is given by the stability matrix $\tilde M_t$
\be
\label{eq:eqno(2.13)}
\delta \ba_t = \tilde M_t\, \delta\ba_0\,.
\ee
The stability matrix $\tilde M_t$ obeys the differential equation
\be
\label{eq:eqno(2.14)}
\frac{\rd }{\rd t} \tilde M_t = \tilde J
\,{\tilde K_t}\, \tilde M_t
\,,
\ee
where we have defined the matrix
$\tilde K_t$ with matrix elements
\be
\label{eq:eqno(2.15)}
\{\tilde K_t\}_{ij} \equiv \frac{\partial^2 H}{\partial \alpha_i\partial
\alpha_j}\;\mbox{\rule[-4mm]{0.25pt}{11mm}
\raisebox{-3mm}{\hspace{3pt}$\ba = \ba_t$}}
\,.
\ee
\section{Semiclassical dynamics in phase space}
\label{sec:sec3}
We will make use of a representation of
quantum dynamics in phase space
and will evaluate traces such as
the trace
$\mbox{Tr}\,[\hat U_t]$ 
of the time evolution operator
$\hat{U}_t$
in a coherent-state basis. 
\subsection{Coherent states and Weyl-Heisenberg operators}
Following
the notation established in \cite{lit86} we denote
{\em coherent states} by
\be
\label{eq:eqno(3.1)}
|\ba_0\rangle = \hat \T(\ba_0)\,|0\rangle
\ee
where the state $|0\rangle$ is the
ground state of a harmonic oscillator
and the {\em Weyl-Heisenberg operator}  $\hat \T(\ba_0)$ 
is  given by
\be
\label{eq:eqno(3.2)}
\hat \T(\ba_0) = {\rm e}^{\displaystyle -{\rm i}(\ba_0 \wedge \hba)/\hbar}
\ee
where $\hba = (\hat \bq,\hat \bp)$. 
Throughout we define 
$\ba_0\wedge \ba_1 = \bq_0.\bp_1-\bq_1.\bp_0$, where $\bq.\bp$ denotes 
the usual scalar product. 
Weyl-Heisenberg operators have the following properties
(all of which may be verified using the Baker-Cambbell-Hausdorff formula
\cite{mer70}). 
\begin{enumerate}
\item{} {\em Translation}
\be
\label{eq:eqno(3.3)}
\hat \T^\dagger(\ba_0)\, \hat\ba \,\hat \T(\ba_0) = \hat\ba + \ba_0\,.
\ee
\item{}
The Weyl-Heisenberg operators satisfy the
following {\em concatenation} property
\be
\label{eq:eqno(3.4)}
\hat \T(\ba_0) \hat \T(\ba_1) = 
{\rm e}^{\displaystyle -{\rm i} 
[\ba_0\wedge \ba_1]/2\hbar} \,\hat \T(\ba_0 + \ba_1)
\,.
\ee
From this it follows that 
$ \hat \T^{-1}(\ba_0) = \hat \T(-\ba_0) = 
\hat \T^\dagger(\ba_0)$.
\item{} {\em Equation of motion}
\be
\label{eq:eqno(3.5)}
{\rm i}\hbar \frac{\rd}{\rd t} \hat \T(\ba_t)
= \hat \T(\ba_t)\,\left(\textstyle{1\over 2} [\dot{\ba}_t \wedge\ba_t]
+[\dot{\ba}_t \wedge \hat\ba]\right)\,.
\ee
\end{enumerate}
The Weyl-Heisenberg operators translate wave
packets in phase space, by virtue of (\ref{eq:eqno(3.3)}).
It is expected that wave packets are not only
moved around but also deformed under quantum evolution. 
In order to represent deformations of
wave packets one needs a representation
of symplectic transformations $\tilde S$
on the Hilbert space of the system.
\subsection{Representation of symplectic transformations}
We seek operators which provide a representation of
symplectic matrices $\tilde S$ on the Hilbert space of the
system. These are termed
{\em metaplectic operators}, and are defined by the 
requirement that the operators $\meta(\tilde S)$ which 
which represent the symplectic matrices $\tilde S$ are unitary
matrices, satisfying
\be
\label{eq:eqno(3.6)}
\meta(\tilde S)\, \hat \T(\ba)\, \meta^{-1}(\tilde S)
= \hat \T(\tilde S\ba)\,.
\ee
Usually \cite{com97,lit86} this representation is
derived in configuration space and explicitly defined 
by prescribing the configuration-space matrix elements
$\langle \bq^\prime | \meta(\tilde S) | \bq\rangle$
of $\meta(\tilde S)$. 
In the mathematical literature so-called
Fock-space representations of symplectic
transformations are often used. These
go back to and are defined in \cite{bar61}. 
Such representations introduce additional
undesirable complications in  the present context.
In the following we use the Weyl representation \cite{gro46}
of the metaplectic operators $\meta(\tilde S)$, in
which an operator is expressed in terms of an
integral over the Weyl-Heisenberg operators. 
This representation is simpler, and also more
natural because it uses phase space throughout,
avoiding making an arbitrary choice of coordinates. 
The only place where we are aware of the Weyl representation
of the metaplectic operators having been constructed 
previously is in \cite{wil98} (the results differ slightly:
the requirement of unitarity was not imposed, the expansion
was in terms of magnetic translations rather than Weyl-Heisenberg
translations, and a phase factor
was included which always evaluates to unity). 

The operator
\be
\label{eq:eqno(3.7)}
\meta(\tilde S) = 
{\exp({\rm i}\pi \nu /2)
\over{\sqrt{\vert {\rm det}(\tilde S-\tilde I)\vert}}}
\int\! \frac{\rd\by}{(2\pi\hbar)^d} \,
\exp\Big(\frac{{\rm i}}{2\hbar} \by.\tilde A \by\Big)\, \hat \T(\by)
\ee
is the metaplectic representative of
the $d$-dimensional symplectic matrix  $\tilde S$ when
the symmetric matrix $\tilde A$ is given by
\be
\label{eq:eqno(3.8)}
\tilde A = \textstyle{{1\over 2}} \tilde J (\tilde S + \tilde I) 
(\tilde S -\tilde I)^{-1}\,.
\ee
The integer index $\nu $ will be discussed further in due course,
but we remark that when all of the eigenvalues of $\tilde A$ are 
large and positive, setting $\nu =d$ makes $\meta (\tilde S)$
approximate the identity operator, $\hat I$.
Following the approach of reference \cite{wil98}, using (\ref{eq:eqno(3.4)}) 
it is easily verified that 
this expression satisfies (\ref{eq:eqno(3.6)}), and that 
the following properties hold:
\begin{enumerate}
\item{}  The operator $\meta(\tilde S)$
represents symplectic transformations
on the quantum-mechanical Hilbert space, in the sense that it 
satisfies (\ref{eq:eqno(3.6)}), and also
\be
\label{eq:eqno(3.9)}
\meta(\tilde S) \, \hba\, \meta^{-1}(\tilde S) 
=\tilde S^{-1}\hba\,.
\ee
\item{} The operator $\meta(\tilde S)$ is {\em unitary}
\be
\label{eq:eqno(3.10)}
\meta(\tilde S) \meta^\dagger(\tilde S) = \hat I\,.
\ee
\item{} The operator $\meta(\tilde S)$ obeys
the following law of {\em concatenation}
\be
\label{eq:eqno(3.11)}
\meta(\tilde S_1) \meta(\tilde S_2) = \pm\,\meta(\tilde S_1 \tilde S_2)\,.
\ee
The origin of the sign will be discussed in due course.
\end{enumerate}
Thus (\ref{eq:eqno(3.7)}) constitutes
the desired quantum-mechanical representation of symplectic 
transformations. 
Furthermore, the following properties
will be used below:
\begin{enumerate}
\addtocounter{enumi}{3}
\item{} The representation of a symplectic matrix with an infinitesimal
generator, $\tilde S = \tilde I\! - \!\epsilon \tilde H \tilde J$ 
(with $\tilde H$ symmetric) is given by
\be
\label{eq:eqno(3.12)}
\meta(\tilde S )  
= \hat I- \frac{{\rm i}\epsilon}{2\hbar} \hba . \tilde J \tilde H \tilde J 
\hba + O(\epsilon^2)\,,
\ee
compare \cite{gro46,deg72}.
\item{} Using the infinitesimal representation (\ref{eq:eqno(3.12)}),
it was shown in \cite{lit86} that the operator $\meta(\tilde S_t)$ obeys
the equation of motion
\be
\label{eq:eqno(3.13)}
{\rm i}\hbar \frac{\rd}{\rd t} \meta(\tilde S_t)
= -\textstyle{1\over 2} \hat{\ba}.\tilde J 
\displaystyle{{\rd \tilde S_t\over{\rd t}}}
\tilde S_t^{-1}\hat{\ba}\, \meta(\tilde S_t)\,.
\ee
\item{} Finally, the {\em trace} of
$\meta(\tilde S)$ is
\be
\label{eq:eqno(3.14)}
\mbox{Tr}\, [\meta(\tilde S)] = 
\frac{\exp({\rm i}\pi \nu /2)}
{\displaystyle\sqrt{\vert \det(\tilde S-\tilde I)\vert }}\,.
\ee
This is immediately obvious from (\ref{eq:eqno(3.7)}) 
by noting that $\mbox{Tr}\,[\hat \T(\by)] = (2\pi\hbar)^d \,\delta(\by)$.
\end{enumerate}

Obviously, the representation
(\ref{eq:eqno(3.7)}) breaks down
whenever $\tilde S$ has eigenvalues
unity. 
It will be seen below that
in a semiclassical description
of wave-packet dynamics, $\tilde S$ is
just the stability matrix of the
classical nearby-orbit problem.
In the case of chaotic Hamiltonian flows, 
this stability matrix always has at least two eigenvalues
unity, corresponding to the direction along
the flow ({\em parallel direction}). 
In this case it is possible,
as will be shown below, to separate  the
metaplectic representations corresponding to
motion in the parallel and perpendicular
directions. For the metaplectics in the
parallel direction, the following representation
will be useful:
the singular $2\times 2$ symplectic matrix
\be
\label{eq:eqno(3.15)}
\tilde S = \left( \begin{array}{cc} 1 & b \\ 0 & 1 \end{array} \right)
\ee
is represented by the metaplectic operator
\be
\label{eq:eqno(3.16)}
\meta(\tilde S) = \frac{1}{\sqrt{2\pi {\rm i}\hbar b}}\,
\int_{-\infty}^\infty 
\rd x_1 \,{\rm e}^{{\rm i} x_1^2/2\hbar b}\,\hat \T(\bx)\\
\ee
with $\bx = (x_1,0)$ (c.f. \cite{fol89,wil98}).
An infinitesimal negative imaginary part added to the real number $b$
makes this integral convergent. Note that the phase of the prefactor
increases by $\pi/2$ when $b$ goes negative.
This representation of a subgroup of symplectic transformations
satisfies the above properties 1 -- 6.
\subsection{Topology of the symplectic group}
\label{sec:sec3.3}
We must now consider the topology of the group of symplectic
matrices and of the corresponding metaplectic operators.
In particular, we must explain the fact that the metaplectic
operators form a \lq double covering' of the symplectic matrices,
since this property gives contributions to the 
{\em Maslov indices} \cite{mas81}
which occur in the Gutzwiller trace formula.

We commence by discussing some results which are already known
\cite{lit86}. 
We consider only the case of $2\times 2$ matrices; the results
generalise to higher dimensions. A general symplectic matrix
may be written in the form
\be 
\label{eq:eqno(3.17)}
\tilde S=\tilde S_{\rm T} \tilde S_{\rm R}
\ee
where $\tilde S_{\rm T}$ and $\tilde S_{\rm R}$ are, 
respectively, a symmetric
matrix satisfying ${\rm det}(\tilde S_{\rm T})=1$, and an orthogonal
matrix. In the case of $2\times 2$ matrices, the matrix $\tilde S_{\rm T}$
belongs to a two-parameter family with the topology of the plane,
and $\tilde S_{\rm R}$ is a rotation matrix parametrised by an angle $\theta$,
so that the space of the orthogonal matrices therefore has a circular 
topology.
The space of $2\times 2$ symplectic matrices therefore has the
topology of the Cartesian product of a plane and a circle. The
symplectic matrices can also be labeled by points in a space
of three real variables,  
as a periodic function of the parameter $\theta$. It will be
helpful to bear both representations in mind. We will term
these two regions of parameter space the \lq fundamental
domain' and the \lq extended domain', respectively. The extended
domain consists of a set of \lq cells', which are replicas
of the fundamental domain generated by adding an integer
multiple of $2\pi $ to $\theta$.

The metaplectic operators $\meta (\tilde S)$ may be represented 
as analytic functions of the underlying symplectic matrices 
$\tilde S$, but this analytic representation could be multi-valued.
We discuss the simplest case, in which the matrix $\tilde S_{\rm T}$ is
the identity matrix, so that $\tilde S$ is a $2\times 2$ rotation
matrix, $\tilde S_{\rm R}(\theta)$. The metaplectic operator representing
a rotation in phase space is clearly generated by the action
of a harmonic oscillator $H(x,p)={1\over 2}(x^2+p^2)$. The 
corresponding metaplectic operator may be written either
in the form (\ref{eq:eqno(3.7)}), or alternatively
\be 
\label{eq:eqno(3.18)}
\meta(\tilde S_{\rm R}(\theta))
=\exp\biggl({-{\rm i}\over {2\hbar}}(\hat x^2+\hat p^2)\theta\biggr)
\,.
\ee
By expanding states in a basis of harmonic oscillator eigenfunctions,
we see that
in the extended domain, this operator clearly satisfies the 
concatenation rule in the form 
$\meta \big(\tilde S_{\rm R}(\theta_1+\theta_2)\big)=
\meta \big(\tilde S_{\rm R}(\theta_1)\big)\,
\meta \big(\tilde S_{\rm R}(\theta_2)\big)$, 
but $\meta \big(\tilde S_{\rm R}(2\pi)\big)=-\hat I$. We therefore see that
in the fundamental domain, the metaplectic representing
rotations is a double covering, in the sense that a product 
of two metaplectics is equal to another metaplectic only
up to a sign. Specifically, for $\theta $ in the fundamental
interval $[0,2\pi)$, we have
\be 
\label{eq:eqno(3.19)}
\meta \big(\tilde S_{\rm R}(\theta_1)\big)\,
\meta \big(\tilde S_{\rm R}(\theta_2)\big)
=(-1)^s\meta \big(\tilde S_{\rm R}((\theta_1+\theta_2)
{\rm mod}2\pi)\big)\ ,\ \ \ 
s={\rm int}\biggl({\theta_1+\theta_2\over {2\pi}}\biggr)
\,.
\ee
This property is general, in that the metaplectic
operators do form a double covering of the symplectic operators
in the fundamental domain, although they can be represented as a 
single valued analytic function in the extended domain. 
This assertion can be checked by means of the following
argument. Firstly, using the properties of the Weyl-Heisenberg
operators, it is verified that (\ref{eq:eqno(3.7)}) satisfies
(\ref{eq:eqno(3.11)}), with a factor $(-1)^s$
appearing rather than some more general phase factor. Continuity
considerations show that if (\ref{eq:eqno(3.19)}) holds when
$\tilde S=\tilde S_{\rm R}$, it must also hold in the more general
case, implying a double valued rather than single valued function
throughout the fundamental domain. The fundamental
domain could be delimited by the planes $\theta=0$ and $\theta=2\pi$
in the representation (\ref{eq:eqno(3.17)}), although there
is an arbitrariness in the choice of the boundaries, and
we will discuss an alternative choice shortly.

We now discuss the index $\nu $ appearing in our general 
representation of the metaplectic operators,
defined by (\ref{eq:eqno(3.7)}). This is an explicit analytic
representation, which is defined everywhere except upon
the  manifold of codimension 1 where ${\rm det}(\tilde S-\tilde I)=0$.
(Alternative representations, such as (\ref{eq:eqno(3.16)}),
are available on this manifold.) The index $\nu$ is chosen so that the 
resulting representation of the metaplectic operators is
continuous across manifolds where ${\rm det}(\tilde S-\tilde I)=0$.
It changes by $\pm 1$ on each crossing. 
The manifolds where ${\rm det}(\tilde S-\tilde I)=0$ are of two
types; the function ${\rm det}(\tilde S-\tilde I)$ is either
increasing or decreasing as $\theta$ increases. These form
an alternating sequence as $\theta $ increases (although they
may coalesce to a double zero, as happens when $\tilde S_{\rm T}=\tilde I$). 
For our purposes it will be convenient to define the domain boundaries
as being one of these two sets of surfaces satisfying the equation 
${\rm det}(\tilde S-\tilde I)=0$.
There are two possibilities for the change of the index upon
crossing to a successive cell. Either the changes of index
have the same sign on each of the two manifolds where 
${\rm det}(\tilde S-\tilde I)=0$, or they have different signs. 
The fact that the metaplectic operators change sign when
$\theta $ increases by $2\pi $ implies that the former 
possibility is realised. The index
$\nu $ therefore changes by $\pm 2$ when crossing to an equivalent
point in an adjacent cell.
\subsection{Semiclassical dynamics}
In Ref. \cite{lit86} the following ansatz for the quantum-mechanical
time evolution operator is discussed for the evolution of the coherent state 
$\vert \ba \rangle$
\be
\label{eq:eqno(3.20)}
|\ba_t\rangle
\equiv \hat U_t |\ba_0\rangle
\sim {\rm e}^{\displaystyle {\rm i}\gamma_t/\hbar}\, 
\hat \T(\ba_t)\,\meta (\tilde S_t) \,|0\rangle\,.
\ee
It is not possible to write a corresponding semiclassical ansatz 
for the operator $\hat U_t$ itself, in a form which is independent
of the label of the coherent state upon which it acts. Nevertheless,
(\ref{eq:eqno(3.20)}) is sufficient for our purposes.
The phase $\gamma_t$ and the matrix
$\tilde S_t$ are to be determined by substitution
into the Schr\"odinger equation
\be
\label{eq:eqno(3.21)}
{\rm i}\hbar \ddt |\ba_t\rangle = \hat H |\ba_t\rangle\,.
\ee
The lhs of (\ref{eq:eqno(3.21)}) gives
\begin{eqnarray}
\label{eq:eqno(3.22)}
&&-\dot\gamma_t \,
{\rm e}^{\displaystyle {\rm i}\gamma_t/\hbar}
\,\hat \T(\ba_t)\,\meta(\tilde S_t)\vac
\nonumber\\
&&+{\rm e}^{\ds {\rm i}\gamma_t/\hbar}\, 
\hat \T(\ba_t)
\,\left(\textstyle {1\over 2} [\dot{\ba}_t \wedge\ba_t]
+[\dot{\ba}_t \wedge \hat\ba]\right)\,
\meta(\tilde S_t)\, \vac
\nonumber\\
&&-\textstyle{1\over 2}
\,{\rm e}^{\displaystyle {\rm i}\gamma_t/\hbar}\,\hat \T(\ba_t)\,
\meta(\tilde S_t)\, \hat{\ba}.\tilde J
\displaystyle{{\rd \tilde S_t\over{\rd t}}}\tilde S_t^{-1}\hat{\ba}\, 
\meta(\tilde S_t)\,\vac\,.
\end{eqnarray}
The rhs of (\ref{eq:eqno(3.21)}) is evaluated using 
\begin{eqnarray}
\label{eq:eqno(3.23)}
&&   \hat H |\ba_t\rangle =  {\rm e}^{\ds {\rm i}\gamma_t/\hbar}\\
\mbox{}\hspace*{1cm}&&\times\left(H(\ba_t;t)  \hat \T(\ba_t) \meta(\tilde S_t)|0\rangle
    + \hat \T(\ba_t) \frac{\partial H}{\partial \ba_t} \hba
      \meta(\tilde S_t)|0\rangle\right .\nonumber \\
\mbox{}\hspace*{1cm}   &&+\left . \textstyle{1\over 2} \hat \T(\ba_t) \hba . \tilde J\,\tilde K_t\, \hba
   \meta(\tilde S_t)|0\rangle + \cdots\right)
\nonumber
\end{eqnarray}
with $\tilde K_t$ given by Eq. (\ref{eq:eqno(2.15)}).
This expression can be obtained from the Weyl quantisation scheme.
One thus obtains the following set of equations,
\begin{eqnarray}
\label{eq:eqno(3.24)}
\dot \gamma_t &=& \textstyle{1\over 2} [\ba_t \wedge \dot\ba_t] -H(\ba_t)\,,\\
\label{eq:eqno(3.25)}
\dot\ba_t &=& \tilde J \,\frac{\partial H}{\partial \ba_t}\,,\\
\label{eq:eqno(3.26)}
{\rd \over {\rd t}} \tilde S_t &=&
 \tilde J\, \tilde K_t \tilde S_t\,.
\end{eqnarray}
From (\ref{eq:eqno(3.24)}) it follows that $\gamma_t$ is
related to Hamilton's action $R_t$
\be
\label{eq:eqno(3.27)}
\gamma_t = R_t - \textstyle{1\over 2}{\bp}_t.{\bq}_t 
+ \textstyle{1\over 2}{\bp}_0.{\bq}_0\ ,\ \ \ R_t=\int {\bp}.d{\bq}-Hdt
\ee
Eqs. (\ref{eq:eqno(3.25)}) are just Hamilton's equations, 
(\ref{eq:eqno(2.11)}). By comparison of (\ref{eq:eqno(2.14)}) and
(\ref{eq:eqno(3.26)}) it follows that $\tilde S_t$ is the stability 
matrix $\tilde M_t$.
The final result is \cite{lit86}
\be
\label{eq:eqno(3.28)}
|\ba_t\rangle
\sim {\rm e}^{\ds {\rm i}(R_t-\bp_t.\bq_t/2+ \bp_0.\bq_0/2)/\hbar}\, 
\hat \T(\ba_t)\,\meta (\tilde M_t) \,\vac    \,.
\ee
The propagation of a state $|\ba_0\rangle$ is thus
envisaged as follows:
$|\ba_0\rangle$ is transported to the origin
of phase space by $\hat \T^\dagger(\ba_0)$, there
it is deformed by $\meta(\tilde M_t)$ and finally
the state is transported to $|\ba_t\rangle$
by $\hat \T(\ba_t)$.

The evolution of the wave packet should be continuous.
The metaplectic operators can be presented as an analytic
function of the symplectic matrices in the extended domain. 
If we follow the evolution of the symplectic 
matrix as the trajectory evolves, the metaplectic operator
evolves continuously, and expression (\ref{eq:eqno(3.28)})
remains valid. The index $\nu $ changes by $\pm 1$ each
time $\tilde M_t$ crosses the a manifold where 
${\rm det}(\tilde M_t-\tilde I)=0$. 
These changes of index are the origin of the Maslov
indices \cite{mas81} which appear in the Gutzwiller trace formula.
\section{Trace formulae for classically chaotic (hyperbolic)
systems}
\label{sec:sec4}
\subsection{Gutzwiller's trace formula for chaotic maps}
\label{sec:sec4.1}
\begin{figure}
\centerline{\psfig{file=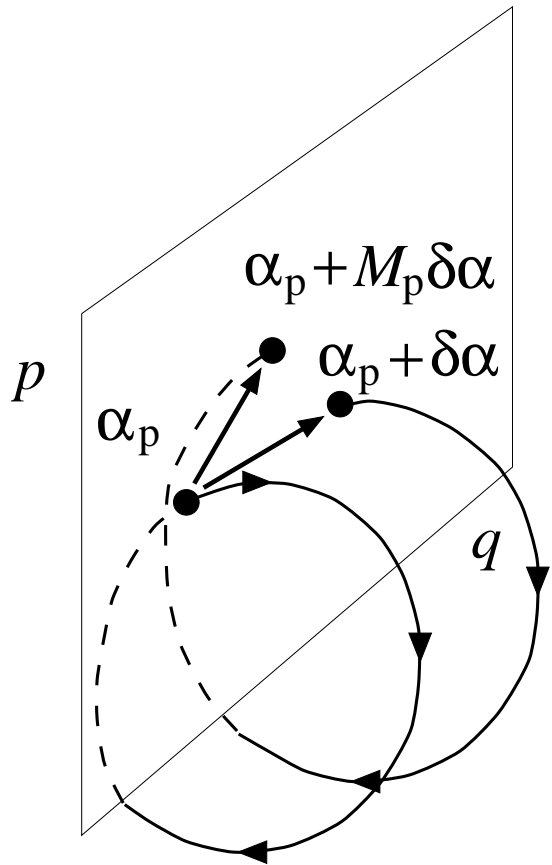,width=7cm,angle=270}}
\mbox{}\\
\caption{\label{fig:nearby1} Illustrates the 
nearby orbit problem for a classical map. Shown
is a periodic orbit $\ba_{\rm p}$ and a nearby orbit
starting at $\ba_{\rm p} + \delta \ba$ and leading
to $\ba_{\rm p} + \tilde M_{\rm p} \delta \ba$ in time $t$.}
\end{figure}

In this section, we consider a time-dependent Hamiltonian
flow with $d=1$  degree of freedom, 
defined by the Hamilton function $H(q,p;t)$,
such as a driven pendulum with a periodic driving
force of period $T$.
We define a stroboscopic
section at $t = n T$, with $n = 0,1, 2,\ldots$.
The dynamics on the stroboscopic section,
generated by the flow $H(q,p;t)$, is given
by a Poincar\'e map  acting on
the stroboscopic section. The quantum evolution
operator of this Hamiltonian, $\hat U_t$ evaluated 
at the stroboscopic time $t=T$ is termed the 
Floquet operator. We will obtain an expression
for the trace of powers of the Floquet operator
$\hat U_{nT}=[\hat U_T]^n$, and use this to give 
a trace formula for the density of eigenphases.
We may write
\be
\label{eq:eqno(4.1)}
\mbox{Tr}\,[ \hat U_t]  = \int\frac{\rd\ba}{2\pi\hbar}\,\langle\ba| 
\hat U_t |\ba\rangle\,.
\ee
Now
\be
\label{eq:eqno(4.2)}
\langle\ba| \hat U_t |\ba\rangle
\sim \exp[{\rm i}(R_t-p_tq_t/2+ p_0q_0/2)/\hbar]\, 
\langle 0| \hat \T^\dagger(\ba)\,\hat \T(\ba_t)\,\meta(\tilde M_t)\vac\,.
\ee
(Here we consider $\meta(\tilde M_t)$ in the extended domain.)
We assume that the classical dynamics is hyperbolic.
In this case, within the semiclassical approximation, 
the essential contributions to this trace will come from isolated,
unstable periodic points $\ba_{\rm p}$
and we write (compare Fig. \ref{fig:nearby1})
\be
\label{eq:eqno(4.3)}
\ba_0 = \ba_{\rm p} + \delta \ba\,,\qquad\ba_t 
= \ba_{\rm p} + \tilde M_{\rm p}\,\delta \ba\,,
\ee
where $t = nT$, and $\tilde M_{\rm p}$ is the stability
matrix evaluated at the periodic point.  
Using (\ref{eq:eqno(3.4)}) and (\ref{eq:eqno(3.6)}) one
has for the matrix element
in the vicinity of $\ba_{\rm p}$ 
\begin{eqnarray}
\label{eq:eqno(4.4)}
&&
\langle 0| \hat \T^\dagger(\ba_{\rm p}+\delta \ba)\,
\hat \T(\ba_{\rm p}+\tilde M_{\rm p}\delta \ba)\,\meta(\tilde M_{\rm p})\vac
\sim 
\exp[{\rm i}\ba_{\rm p} \wedge (\tilde M_{\rm p}-\tilde I)\delta \ba/2\hbar]
\\
&&\hspace*{7cm}\times \langle\delta \ba| 
\meta(\tilde M_{\rm p}) |\delta \ba\rangle\,.
\nonumber
\end{eqnarray}
Consider the expansion of the phase in (\ref{eq:eqno(4.2)})
about the phase value $R_{\rm p}$ characterising the period orbit.
From (\ref{eq:eqno(3.24)}), the difference $\delta \gamma_t$
between the phases characterising two nearby trajectories 
satisfies the following differential equation:
\be
\label{eq:eqno(4.5)}
\delta \dot \gamma_t=-{\textstyle{1\over 2}}\delta \ba_t \wedge \dot \ba_t
-{\textstyle{1\over 2}}\ba_t \wedge \delta \dot \ba_t
-{\textstyle{1\over 2}}\delta \ba_t \wedge \delta \dot \ba_t
-\delta H
=-{\textstyle{1\over 2}}{\rd\over \rd t}
\bigl(\ba_t \wedge  \tilde M_t \delta \ba \bigr)\,.
\ee
This equation must be solved subject to the boundary
condition that $\delta \gamma_0=0$. The solution is
\be
\label{eq:eqno(4.6)}
\delta \gamma_t=-{\textstyle{1\over 2}}\ba_t\wedge \tilde M_t \delta \ba
+{\textstyle{1\over 2}} \ba_0\wedge\delta \ba
\ee
For a periodic orbit, $\ba_t=\ba_0=\ba_{\rm p}$, so that in the neighbourhood
of such an orbit (with $\tilde M_{\rm p}= \tilde M_t$)
\be
\label{eq:eqno(4.7)}
\delta \gamma=-{\textstyle{1\over 2}} 
\ba_{\rm p}\wedge (\tilde M_{\rm p}-\tilde I)\delta \ba\,.
\ee
Expanding the phase in (\ref{eq:eqno(4.2)})
in the vicinity of $\ba_{\rm p}$, one now obtains
\be
\label{eq:eqno(4.8)}
R_t -p_t q_t/2+p_0q_0/2
= R_{\rm p}-\textstyle{1\over 2}
[\ba_{\rm p}\wedge(\tilde M_{\rm p}-\tilde I)\delta \ba]
+O(\delta \ba^3)
\,.
\ee
Taking everything together
(note that all terms linear in $\delta \ba$ in the phase cancel)
we obtain for the contribution ${\rm Tr}[\hat U_t]_{\rm p}$ 
of a periodic point $\ba_{\rm p}$ to the trace of $\hat U_t$:  
\be
\label{eq:eqno(4.9)}
{\rm Tr}[\hat U_t]_{\rm p}\sim  
\exp({\rm i}R_{\rm p}/\hbar) \,\mbox{Tr}\,[\meta(\tilde M_{\rm p})]\,.
\ee
It thus turns out that the stability property of
the periodic orbit is encoded in the 
trace of $\meta(\tilde M_{\rm p})$.  Using (\ref{eq:eqno(3.14)}) we have
\be
\label{q:eqno(4.10)}
{\rm Tr}[\hat U_t]_{\rm p}\sim 
 \frac{\exp({\rm i}R_{\rm p}/\hbar+{\rm i}\pi \nu_{\rm p}/2)}
 {\sqrt{\vert {\rm det}(\tilde M_{\rm p}-\tilde I)\vert}}\,.
\ee
This contribution is summed over all periodic points
and one obtains
\be
\label{eq:eqno(4.11)}
\mbox{Tr}\,[\hat U_T^n]
\sim  
\sum_{\stackrel{\mbox{\small periodic points $p$}}
{\mbox{\small of order $n$ }}}
\frac{n_{p}\,\exp({\rm i}R_{p}/\hbar+{\rm i}\pi \nu_{\rm p}/2)}
{\sqrt{\vert \det(\tilde M_p-\tilde I)\vert}}\,,
\ee
where $p$ is now an integer index labeling the periodic points.
The sum is over all periodic points,
including primitive points and their
repetitions. In the latter
case, $n_p = n/r_p$ where $r_p$ is
the order of repetition.
With this result one finds for the density (\ref{eq:eqno(2.10)})
of eigenphases
\be
\label{eq:eqno(4.12)}
 d(\theta)\sim  \frac{N}{2\pi\hbar} - \frac{1}{\pi\hbar}\mbox{Im}
 \sum_{p}
\frac{n_p\, \exp[{\rm i}(R_p-n_p r_p\theta)/\hbar  
+ {\rm i}\pi\nu_{p}/2]}
 {\sqrt{|\det(\tilde M_p-\tilde I)|}}\,.
 \ee
This is Gutzwiller's trace formula for chaotic maps, discussed
in reference \cite{smi94}.
The first term is often called Weyl contribution and
describes the average density of states. 
\subsection{Gutzwiller's trace formula for chaotic Hamiltonian flows}
\label{sec:sec4.2}
In this section we consider chaotic Hamiltonian flows
with $d=2$ degrees of freedom and a time-independent
Hamilton function $H(\bq,\bp)$. 
As in the previous section,
a semiclassical approximation for the trace
of the quantum-mechanical time evolution operator $\hat U_t$ 
will be derived,
in order to find a semiclassical expression
for the density of states.
The trace $\mbox{Tr}\,[\hat U_t]$ will again be evaluated using coherent
states, using (\ref{eq:eqno(1.2)}). 
There are two contributions: a smooth part
(Weyl term) due to very short orbits (corresponding to the first term
in (\ref{eq:eqno(4.12)}))  and oscillatory contributions
from periodic orbits. 

We consider the smooth part first.
For very short times,
\be
\label{eq:eqno(5.1)}
\langle\ba| \hat U_t |\ba\rangle
\simeq 
\langle\ba |\hat \T(\delta \ba_t) \vert \ba\rangle \,
{\rm e}^{\ds -\frac{{\rm i}}{\hbar} H(\ba)t}
\ee
where $\delta \ba_t=\ba_t-\ba$. If the coherent
states are those of the Hamiltonian $H(\ba )={1\over 2}\ba.\ba$,
then we may write
\be 
\label{eq:eqno(5.2)}
\langle\ba |\hat \T(\delta \ba_t) \vert \ba\rangle \,
=f(\vert \delta \ba_t\vert/\hbar)
\ee
where $\vert \ba \vert$ is the Euclidean norm of $\ba$, (and the
function $f(x)$ is a Gaussian, satisfying $f(0)=1$).
Thus the smooth part of the density is
\begin{eqnarray} 
\label{eq:eqno(5.3)}
\langle \rho (E) \rangle & \sim &
\int \!\frac{\rd\ba}{(2\pi \hbar)^2}\, 
\int_{-\infty}^\infty \frac{\rd t}{2\pi\hbar}\,
 {\rm e}^{\ds{\rm i}[E-H(\ba)]t/\hbar}
f(\vert \delta \dot \ba \vert t/\hbar)
\nonumber\\
&=&
\int \frac{\rd\ba }{(2\pi \hbar)^2}\,
{1\over {\vert \dot \ba_t\vert \sqrt \hbar}}
\widetilde f[(E-H(\ba)/\vert \delta \dot \ba_t\vert\sqrt\hbar]
\nonumber\\
\end{eqnarray}
where $\widetilde f(y)$ is the Fourier transform of $f(x)$. 
Now, noting that
\be 
\label{eq:eqno(5.4)}
\lim_{\epsilon \to 0}{1\over \epsilon}\tilde f(x/\epsilon)
=f(0)\delta (x)
\,,
\ee
the contribution to the density of states from short times is
\be
\label{eq:eqno(5.5)}
\langle \rho(E) \rangle 
\sim \int\!\frac{\rd\ba}{(2\pi\hbar)^2} \,\delta [E-H(\ba) ]
\,,
\ee
which is the well known Weyl estimate \cite{gut90}.

We now turn to the periodic-orbit contributions.
The derivation of the previous section needs
to be slightly modified since (as we shall shortly 
show) the stability
matrix for a flow always has an eigenvalue
unity (along the direction of the flow). 
Thus the representation (\ref{eq:eqno(3.7)})
is singular. The solution to this problem
is to treat longitudinal and transverse
coordinates separately, as described
in the following.

It will be helpful to introduce a local canonical coordinate
system in the vicinity of an unstable periodic orbit. The reference
orbit is denoted by $\ba_{\rm p}(\tau,\eta)$, with
$\tau$ measuring the time taken a reach the point $\ba_{\rm p}(\tau,\eta)$
from some arbitrary starting point on the orbit at energy $\eta$. 
The period of the orbit is $T_{\rm p}(\eta)$. Any point close to the orbit
may be specified by giving its deviation from some
point on the reference orbit: $\ba= \ba_{\rm p}(\tau,\eta)+\delta \ba$,
where $\delta \ba$ depends upon $\tau$ and $\eta$. 
For points close to the orbit at a reference energy $\eta_0$, 
we will find it convenient
to represent the deviations by a transformed set of local
canonical coordinates, $\delta \ba'$, such that 
\be
\label{eq:eqno(5.6)}
\delta \ba=\tilde M' (\tau) \delta \ba'\ ,\ \ \ 
\delta \ba'=(\delta \tau, \delta E, \delta X, \delta P)
\,,
\ee
where $\tilde M'$ is a 
symplectic matrix depending upon $\tau$. This matrix is chosen
so that the new coordinates $\delta \ba$ 
have the following interpretations. The energy $\delta E$ is 
defined by $\delta E=H(\ba)-\eta_0$. The time coordinate 
$\delta \tau$ measures the contribution to the displacement
$\delta \ba $ from a vector parallel to the flow, namely
$\tilde J\partial H/\partial \ba$. The remaining coordinates
$\delta \bb=(\delta X,\delta P)$ measure perpendicular 
displacements from the orbit; they are chosen such that
$\delta \bb=0$ corresponds to a point on the reference
periodic orbit at energy $\eta_0+\delta E$. The construction
of the symplectic transformation $\tilde M'(\tau)$ is described
in an appendix. Note that this symplectic transformation
is periodic: $\tilde M'(\tau+T_{\rm p}(\eta_0))=\tilde M'(\tau)$.

Given an initial deviation $\delta \ba$ from the reference 
trajectory $\ba_{\rm p}(\tau ,\eta_0)$ at time $t=0$, there will be 
a corresponding deviation $\ba_t$ from $\ba_{\rm p}(\tau +t,\eta_0)$ 
at a later time $t$. We write 
\be
\label{eq:eqno(5.7)}
\delta \ba _t=\tilde M'(t+\tau)\delta \ba'_t\ ,\ \ \ 
\delta \ba'_t=\tilde M''_t \delta \ba'
\,,
\ee
where $\tilde M''_t$ is a monodromy matrix which describes 
the instability of the orbit in the local canonical coordinates
$\delta \ba'$. The construction
of the $\delta \ba'$ coordinates implies that there is no
coupling between the deviation in the $(\delta E,\delta \tau)$
subspace and the delta $\delta \bb$ subspace.
\be
\label{eq:eqno(5.8)}
\delta \ba' = \bigl(\delta\tau,\delta\eta,\delta \bb\bigr)\,,\qquad
\delta \ba'_t = \bigl(\delta\tau+\frac{\partial t}{\partial E}\delta\eta,
\delta\eta,\tilde M_\perp\delta \bb\bigr)
\,,
\ee
so that the stability 
matrix $\tilde M''_t$ has the following form
\be
\label{eq:eqno(5.9)}
\tilde M''_t
= 
\left(
\begin{array}{cc}
 \tilde M_{||} & \tilde 0 \\
\tilde 0 &  \tilde M_{\perp}  
\end{array}
\right)
\ee
where $\tilde M_нн$ and $\tilde M_\perp$ are $2\times 2$ symplectic
matrices, with
\be
\label{eq:eqno(5.10)}
\tilde M_{||}
= 
\left(
\begin{array}{cc}
 1        & b \\
0 &  1            
\end{array}
\right)\,,
\ee
and $b = {\rm d}  t/{\rm d} E$, and $\tilde M_{\perp}$ is a 
$2(d-1)\times 2(d-1)$ symplectic matrix 
mapping infinitesimal deviations
in the transverse coordinates.
In the fixed canonical coordinate system, the stability matrix describing
the evolution of small deviations $\delta \ba$ from the reference 
trajectory is 
$\tilde M_t=\tilde M'(t+\tau)\tilde M''_t[\tilde M'(\tau)]^{-1}$.

Evaluation of the trace ${\rm Tr}[\hat U_t]$ follows the approach
used in section \ref{sec:sec4.1}, and requires the evaluation of 
$\langle \delta \ba \vert \meta (\tilde M_t)\vert \delta \ba \rangle$. 
The metaplectic operator $\meta (\tilde M_t)$ must be considered
a a continuous function of $t$, defined in the extended domain.
It may be written as a product, so that when $t\approx T_{\rm p}(\eta_0)$:
\begin{eqnarray} 
\label{eq:eqno(5.11)}
{\rm Tr}\bigl[\meta(\tilde M_t)\bigr]
&=&{\rm Tr}\bigl[\meta(\tilde M'(t+\tau))
\meta(\tilde M''_t)[\meta(\tilde M'(\tau))]^{-1}\bigr]
\nonumber\\
&=&(-1)^s{\rm Tr}\bigl[
\meta(\tilde M'(\tau))\meta(\tilde M_t'')[\meta(\tilde M(\tau))]^{-1}\biggr]
\nonumber\\
&=&(-1)^s{\rm Tr}\bigl[\meta(\tilde M''_t)\bigr]
\,.
\end{eqnarray}
Here we have used the fact that $\tilde M'(\tau)$ is periodic,
so that $\meta(\tilde M'(\tau))$ is periodic apart from a 
factor $(-1)^s$ discussed in section \ref{sec:sec3.3}.

The operator $\meta(\tilde M''_t)$ may be written
in the form
\be
\label{eq:eqno(5.12)}
\meta(\tilde M''_t)=
\left(\begin{array}{cc}
\meta(\tilde M_\perp) & \hat 0 \\
\hat 0 & \meta(\tilde M_{нн})
\end{array}\right)             
\ee
where $\meta(M_{нн})$ and $\meta(M_\perp)$ are operators acting upon
Hilbert spaces corresponding to respectively one and $d-1$
degrees of freedom. These operators may be expressed as integrals
of the form (\ref{eq:eqno(3.7)}), constructed using sets of mutually
commuting translation operators, respectively  $\hat \T(\tau,\eta)$ 
and $\hat \T(\bb)$.
This implies that the matrix element of the metaplectic operator
$\meta (\tilde M''_t)$ is in the form of a product
\be
\label{eq:eqno(5.13)}
\langle \delta \ba|\meta(\tilde M''_t)|\delta \ba\rangle = 
\langle \delta\tau,\delta \eta|\meta(\tilde M_{||})
|\delta\tau,\delta \eta\rangle
\,
\langle \delta \bb|\meta(\tilde M_\perp)|\delta \bb\rangle\,.
\ee
We now evaluate a contribution to ${\rm Tr}[\hat U_t]$ arising 
from a periodic orbit. The dominant contribution arises for period 
orbits of period $T_{\rm p}=t$, and we choose the energy $\eta$ such that
this condition is satisfied. Following the argument
leading from (\ref{eq:eqno(4.2)}) to (\ref{eq:eqno(4.9)}), the contribution
to the trace is
\begin{eqnarray}
\label{eq:eqno(5.14)}
\mbox{Tr} [\hat U_t]_{\rm p}&=&
{(-1)^s\over{(2\pi\hbar)^2}}\int_0^{T_{\rm p}} \rd\tau
\int_{-\infty}^\infty \rd\delta \eta  \int \rd\delta \bb \ 
\langle \ba_{\rm p}+\delta \ba \vert \hat U_t 
\vert \ba_{\rm p}+\delta \ba \rangle
\nonumber\\
&=&
{\exp({\rm i} R_{\rm p}/\hbar+{\rm i}\pi s)\over{2\pi \hbar}}
\int_0^{T_{\rm p}} \!\!\rd\tau\,
\int_{-\infty}^\infty \!\!\rd\delta \eta \ 
\langle \delta \eta \ \vert \meta(\tilde M_{\vert\vert})
\vert \delta \eta \rangle\ 
\mbox{Tr}\,[\meta(\tilde M_\perp)]
\end{eqnarray}
where $R_{\rm p}$ is the action of the periodic orbit, with
the energy $\eta$ chosen such that $T_{\rm p}(\eta)=t$.
Using Eqs. (\ref{eq:eqno(3.14)}) and (\ref{eq:eqno(3.16)}), and 
performing the integral over $\tau $, this evaluates to
\be
\label{eq:eqno(5.15)}
\mbox{Tr} [\hat U_t]_{\rm p}=
{T_{\rm p}\over {\sqrt{2\pi {\rm i} \hbar b_{\rm p}}}}
{\exp[{\rm i}R_{\rm p}/\hbar+{\rm i}\pi (\mu_{\rm p}+2s)/2)]
\over {\sqrt{\vert \det(\tilde M_\perp-\tilde I)\vert_{\rm p}}}}
\,.
\ee
where $\tilde M_\perp$ is evaluated for the periodic orbit with
period $T_{\rm p}=t$, and $\mu_{\rm p}$ is an additional
index chosen so that $\meta(\tilde M_\perp)$
is a continuous function of time along the periodic orbit. This
expression is an explicit function of $\eta$, but implicitly
a function of $t$, since $\eta $ satisfies $t=T_{\rm p}(\eta)$.
 
Now the time integral over $\mbox{Tr} [\hat U_t]$ is performed,
in order to evaluate the density of states. The contribution
due to a single family of periodic orbits is
\be 
\label{eq:eqno(5.16)}
\rho(E)=\int_{-\infty}^\infty \frac{\rd t}{2\pi\hbar}\ 
\exp({\rm i}Et/\hbar)\, {\rm Tr}[\hat U_t]_{\rm p}
\,.
\ee
This integral is approximated using the method of stationary phase.
The time is written $t = t_0 + \delta t$ where $t_0$,
the stationary phase point, satisfies
\be
\label{eq:eqno(5.17)}
\ddt (Et+R_{\rm p}) = 0\qquad \mbox{for $t=t_0$}\,.
\ee
We will discuss the solution of this equation 
in detail. 
For each value of the time
$t$ over which we integrate, the integrand is evaluated
at a different value of the energy, $\eta$, determined
by the implicit relation $t=T_{\rm p}(\eta)$. The action
$S_{\rm p}$ of the orbit and its period $T_{\rm p}$ are also functions
of the energy $\eta $:
\be 
\label{eq:eqno(5.18)}
S_{\rm p}(\eta)=\oint {\bp}.\rd{\bq}
\bigg\vert_{({\bq},{\bp})=\ba_{\rm p}(\tau)}
\ ,\ \ \ \ T_{\rm p}={\partial S_{\rm p}\over {\partial \eta}}\,.
\ee
The equation (\ref{eq:eqno(5.17)}) for the stationary
phase condition now becomes
\begin{eqnarray}
\label{eq:eqno(5.19)}
0&=&{\rd\over {\rd t}}\big\{S_{\rm p}[\eta(t)]-\eta(t)t+Et\big\}
=[T_{\rm p}(\eta)-t]{\rd\eta\over{\rd t}}+[E-\eta(t)]\,.
\end{eqnarray}
We have already required that $T_{\rm p}(\eta)=t$ and we therefore
satisfy the stationary phase condition by requiring, in addition, 
that $E=\eta(t)$. When evaluating the integral (\ref{eq:eqno(5.16)}),
the value of the energy $E$ therefore determines $\eta=E$, which
in turn determines the stationary phase point $t_0=T_{\rm p}(\eta)$.
We note that this corresponds to a Legendre transformation
from $R$ to $S$:
\be
\label{eq:eqno(5.20)}
S_{\rm p} = R_{\rm p} + Et\,,\qquad
\frac{\partial R_{\rm p}}{\partial t} = -E\,.
\ee
In order to perform the stationary phase integral, we require
the second derivative of the action: we write
\be 
\label{eq:eqno(5.21)}
Et+S_{\rm p}(\eta(t))-\eta t=S_{\rm p}(E)+{\textstyle{1\over 2}}
\lambda[t-T_{\rm p}(E)]^2
\ee
where we find
\be
\label{eq:eqno(5.22)}
\lambda
=-{\rd E\over {\rd T_{\rm p}}}=-b^{-1}\vert_{\rm p}\,.
\ee
The Gaussian integration over $\delta t$ gives 
a factor of $\exp({\rm i}\pi/2)\sqrt{2\pi {\rm i}\hbar b_{\rm p}}$, 
and thus we obtain for the 
contribution of a periodic orbit to the density of states
\be
\label{eq:eqno(5.23)}
\frac{1}{2\pi\hbar}
\frac{T_{\rm p}\,\exp\bigl({\rm i}S_{\rm p}/\hbar
+{\rm i}\pi \nu_{\rm p}/2 \bigr)}
{\sqrt{\vert \det(\tilde M_\perp-\tilde I)\vert_{\rm p}}}
\ee 
where $\nu_{\rm p}=\mu_{\rm p}+2s+1$ is the 
Maslov index of the periodic orbit.
By summing over individual contributions from
isolated, hyperbolic periodic orbits, labeled by an index $p$.
Combining contributions from positive- and negative-time orbits
one obtains:
\be
\label{eq:eqno(5.24)}
\rho (E) = \langle \rho(E)\rangle
+ \frac{1}{\pi\hbar} \mbox{Re}
\sum_{p}
{{T_p}\over {\sqrt{|{\rm det}(\tilde M_\perp-\tilde I)|_p}}}
\exp\biggl({{\rm i}S_p\over {\hbar}} + {{\rm i}\pi\nu_{p}\over 2}
\biggr)
\,.
\ee
This is Gutzwiller's trace formula \cite {gut67,gut71,gut90}.
\subsection{More general trace formulae for chaotic Hamiltonian flows}
\label{sec:sec4.3}
The results of the previous section may be used
to derive semiclassical approximations for more general traces, such as
$\mbox{Tr}\,[\hat A\,\hat U_t]$,
where $\hat A$ is an  observable 
with a classical limit $A(\ba)$, compare Eq. (\ref{eq:eqno(2.6)}).
This trace may be written as
\be
\label{eq:eqno(6.1)}
\mbox{Tr}\,\bigl[ \hat A\hat  U_t \bigr] 
= \int\!\frac{\rd\ba}{(2\pi\hbar)^d}\  
A(\ba)\ \langle \ba |\hat U_t|\ba\rangle\,.
\ee
Using (\ref{eq:eqno(5.1)}) one obtains for the smooth
part of the density $\rho_A(E)$ \cite{fei85}
\be
\label{eq:eqno(6.2)}
\langle \rho_A(E) \rangle  
= \int\!\frac{\rd\ba}{(2\pi\hbar)^d} \,A(\ba)\,\delta [E-H(\ba) ]
\,.
\ee
As far as the periodic-orbit contributions are
concerned, it is immediately obvious that all
steps performed in section \ref{sec:sec4.2}
are the same in this case, except
for an additional factor of $A(\tau,E)$\,.
One thus obtains for the weighted density \cite{wil88}
\be
\label{eq:eqno(6.3)}
\rho_A(E)
=
\langle \rho_A(E)\rangle
+ \frac{1}{\pi\hbar} \mbox{Re}
\sum_{p}
\frac{A_p}{\sqrt{|\det({\tilde M_\perp}-\tilde I)|_p}}
\,
\exp\biggl({{\rm i}S_p\over \hbar} + {{\rm i}\pi \nu_p\over 2}
\biggr)
\,.
\ee
where $A_p$ is the average of the observable $A(\ba)$ over the
$p^{\rm th}$ orbit:
\be
\label{eq:eqno(6.4)}
A_p = \int_0^{T_p}\!\rd t\ A(\ba_t)
\,.
\ee
Similarly, a trace formula involving non-diagonal matrix
elements may be derived, starting from
the expression
\be
\label{eq:eqno(6.5)}
S_A(t,s) 
= \mbox{Tr} [\hat A \,\hat U_s\, \hat A\, \hat U_{t-s}]\,.
\ee
We may write
\be
\label{eq:eqno(6.6)}
S_A(t,s) = {1\over{(2\pi \hbar)^d}}
\int\!\frac{\rd\ba}{(2\pi\hbar)^2}\,
A(\ba) \,A(\ba_s) \,\langle \ba | \hat U_t | \ba\rangle\,.
\ee
Using (\ref{eq:eqno(5.1)}), one has \cite{fei85,wil87}
\be
\label{eq:eqno(6.7)}
\langle S_A(E,\hbar\omega)\rangle
= \int_{-\infty}^\infty\!\frac{\rd s}{2\pi\hbar}\,
 {\rm e}^{\ds i\omega s} \int\!\frac{\rd\ba}{(2\pi\hbar)^d}\, 
A(\ba)\,A(\ba_s)\,\delta [E-H(\ba) ]
\,,
\ee
which is just the Fourier transform of
the phase-space averaged correlation
function, as derived in \cite{wil87}. 
Including the periodic-orbit contributions one obtains
the following
expression
\be
\label{eq:eqno(6.8)}
S_A(E,\hbar\omega)=
\langle S_A(E,\hbar\omega)\rangle
+\frac{1}{\pi\hbar} \mbox{Re}
\sum_{p}
\frac{S_p(E,\hbar\omega)}{\sqrt{|\det({\tilde M_\perp}-\tilde I)|_p}}
\,\exp\biggl({{\rm i}S_p\over {\hbar}} + {{\rm i}\pi \nu_p\over 2}
\biggr)
\ee
where 
\be
\label{eq:eqno(6.9)}
S_p(E,\hbar\omega) =  \int_{-\infty}^\infty\!\frac{\rd s}{2\pi\hbar}\,
 {\rm e}^{\ds {\rm i}\omega s}
\int_0^{T_p}\!\rd t\, A(\bq_{t+s},\bp_{t+s})\, A(\ba_t,\bp_t)\,.
\ee
\section{Discussion}
\label{sec:sec5}
We close with a couple of additional remarks.
First, the above formulae were derived
assuming that all periodic orbits are
isolated and unstable (as is the
case in hyperbolic systems). The above approach is very easily extended
to deal with {\em integrable} systems where
the dynamics is most conveniently
expressed in {\em angle} and {\em action} variables
$\ba = (\bth,\bI)$.
In these variables, the Hamilton function
is a function of the action variables only,
$H(\bI)$. Thus
\be
\label{eq:eqno(7.1)}
\bI_t = {\rm const.}\,,\qquad \bth_t = \bth + \bw(\bI)\,t
\ee
with $\bw(\bI) = \partial H/\partial \bI$.
The for the typical case where the frequencies are not
rationally related, the trajectories fill $d$ dimensional
tori in phase space. The periodic orbits only occur when
the frequencies are rationally related, which occurs
when the actions $\bI(\bnu)$ satisfy the implicit equation
\be
\label{eq:eqno(7.2)}
\bw[\bI(\bnu)] t = 2\pi \bnu\,.
\ee
where $\bnu = (\nu_1,\nu_2,\ldots)$ is a vector of integers.
The trajectories labeled by the integer vectors $\bnu$
form $d$-parameter families ({\em rational tori)} which can be labeled
by an initial angle $\bth_0$. It is thus clear that
the leading-order contribution
to the  trace (\ref{eq:eqno(1.2)}) comes from
periodic orbits on rational tori. The derivation of
the trace formula for integrable systems \cite{ber77}
proceeds exactly as before, using a $2d$-dimensional
generalization of the representation (\ref{eq:eqno(3.16)}).

Secondly, it is clear that the trace formulae
Eqs. (\ref{eq:eqno(2.5)}), (\ref{eq:eqno(5.24)}),
(\ref{eq:eqno(6.3)}) and (\ref{eq:eqno(6.8)})
are not exact and at best asymptotically
valid for small $\hbar$. In order to obtain
convergent expressions, the densities 
(\ref{eq:eqno(2.1)}-\ref{eq:eqno(2.3)})
are often smoothed by convolution with
a smoothing factor, i.e.,
\be
\rho_\epsilon(E) = \frac{1}{\sqrt{2\pi}\epsilon}\,
\int_{-\infty}^\infty\!\rd E'\,
\rho(E')\, \exp[-(E-E')^2/2\epsilon^2]\,.
\ee
This leads to a truncation of the periodic
orbit sums in (\ref{eq:eqno(5.24)}),
(\ref{eq:eqno(6.3)}) and (\ref{eq:eqno(6.8)}) \cite{wil87}.
The periodic-orbit formulae (\ref{eq:eqno(5.24)}) and
(\ref{eq:eqno(6.3)}) are certainly
valid provided the wave packet does not spread to
the extent that the use of a linearised approximation 
to describe its deformation becomes invalid. If the system 
is hyperbolic, this criterion leads to the requirement
that $\epsilon > \hbar \lambda/\mbox{log}(S_0/\hbar)$,
where $\lambda $ is an exponent describing the exponential 
separation of trajectories, and $S_0$ a characteristic action
scale for the system.
In the case of (\ref{eq:eqno(6.7)}),
an additional smoothing with respect to the second argument
of $S_A$ is needed, of the same order of magnitude
as $\epsilon$.

\newpage
\mbox{}\hspace*{-5.5mm}{\bf Appendix}\\[0.2cm]
This appendix explains the construction of the 
coordinate system used in section \ref{sec:sec4.2}.
It is constructed around a periodic
orbit. The set of points on the periodic orbit is 
$\ba_{\rm p} (\tau, \eta)$, where $\eta $ is the energy and
$\tau $ is the time taken to reach the point starting
from an arbitrary reference point. We assume that 
these reference points are chosen such that $\ba (\tau, \eta)$
is a smooth function of $\tau$. 

Local canonical
coordinates $(\delta \tau,\delta \eta,\delta X,\delta P)
=(\delta \alpha'_1,\delta \alpha'_2,\delta \alpha'_3,\delta \alpha'_4)
=\delta \ba'$ are constructed from the original coordinates 
$(x,y,p_x,p_y)$ by obtaining 
a set of four vectors ${\bf v}_i$, such
that $\sum_i {\bf v}_i\delta \alpha'_i$ is the variation of $(x,y,p_x,p_y)$ 
resulting from a variation of the new coordinates.
The vector ${\bf v}_1$ is the velocity vector of the Hamiltonian
flow, ${\bf v}_1=\tilde J(\partial H/\partial {\ba})$. 
The vector ${\bf v}_2=\partial \ba_{\rm p} (\tau, \eta)/\partial \eta$
satisfies ${\bf v}_1^{\rm T}\tilde J {\bf v}_2=1$, so that the
coordinates $\delta \alpha_1'=\delta \tau$ and 
$\delta \alpha_2'=\delta \eta$ form a canonical pair.
The remaining vectors ${\bf v}_3$ and ${\bf v}_4$ are 
constructed so that $\gamma_{ij}={\bf v}_i^{\rm T}\tilde J{\bf v}_j$
satisfies $\gamma_{ij}=0$ when $i\in \{1,2\}$ and $j\in\{3,4\}$,
and $\gamma_{34}=1$. These values of $\gamma_{ij}$ represent
five linear constraints on the eight coefficients defining
the vectors ${\bf v}_3$ and ${\bf v}_4$.
All of these vectors are functions of $\tau$. The matrix 
element $M_{ij}'(\tau)$ of $\tilde M'(\tau)$ is the $i^{\rm th}$
component of the vector ${\bf v}_j$.
    \end{document}